\documentclass[12pt]{iopart}

\usepackage{iopams}
\usepackage{graphicx}

\newcommand{\eqref}[1]{(\ref{#1})}

\newcommand{\dd}{\textrm{d}}
\newcommand{\bp}{{\bf p}}
\newcommand{\bmuu}{{\boldsymbol \mu}}
\newcommand{\bnuu}{{\boldsymbol{\nu}}}
\newcommand{\bpt}{{\tilde{\bf p}}}
\newcommand{\pti}{{\tilde{p}}}
\newcommand{\ft}{\tilde{f}}
\newcommand{\obp}{{\bf p}'}
\newcommand{\opt}{{\tilde{p}}'}
\newcommand{\oW}{W'}
\newcommand{\of}{f'}
\newcommand{\oA}{A'}
\newcommand{\ovc}{c'}
\newcommand{\obpt}{{\tilde{\bf p}}'}
\newcommand{\oft}{{\tilde{f}}'}
\newcommand{\ozeta}{{\zeta}'}

\newcommand{\oPi}{\Pi'}
\newcommand{\bo}{\hat{\boldsymbol \omega}}
\newcommand{\bv}{{\bf v}}
\newcommand{\bsi}{\hat{\boldsymbol \sigma}}
\newcommand{\bt}{\hat{\boldsymbol \tau}}

\graphicspath{{Figures/}}

\begin{document}

\title[Initial and final conditions for the fluctuation Relation in Markov processes]{Relevance of initial and final conditions for the Fluctuation Relation in Markov processes.}

\author{A Puglisi$^1$, L Rondoni$^2$ and A Vulpiani$^3$}
\address{$^1$ Dipartimento di Fisica, Universit\`a La Sapienza, p.le Aldo Moro 2, 00185 Roma, Italy}
\address{$^2$ Dipartimento di Matematica and INFM, Politecnico di Torino, Corso Duca degli Abruzzi 24, I-10129, Torino, Italy}
\address{$^3$ Dipartimento di Fisica and INFN, Universit\`a La Sapienza, p.le Aldo Moro 2, 00185 Roma, Italy}

\ead{andrea.puglisi@roma1.infn.it}

\begin{abstract}
Numerical observations on a Markov chain and on the continuous Markov
process performed by a granular tracer show that the ``usual''
fluctuation relation for a given observable is not verified for finite
(but arbitrarily large) times. This suggests that some terms which are
usually expected to be negligible, i.e.\ ``border terms'' dependent
only on initial and final states, in fact cannot be
neglected. Furthermore, the Markov chain and the granular tracer
behave in a quite similar fashion.
\end{abstract}

\maketitle

\section{Introduction}

Non-equilibrium systems are at the center of old but still very lively
research in physics, due to the great importance that out of
equilibrium phenomena have in  the natural
world~\cite{zwanzig,galla,kreuzer}. Since the seminal studies by
Carnot, Clausius and Kelvin, who made clear the relevance of the lack
of equilibrium in the extraction of work from heat, systems out of
thermodynamic equilibrium are investigated in the vastest set of
applications, ranging from mechanical engineering to the study of
chemical reactions, molecular ratchets and all the biophysical
processes of growth, differentiation, movement and evolution.

On the theoretical side, the study of fluctuations in non-equilibrium
systems has been marked by a series of fundamental breakthroughs, such
as Einstein's work~\cite{einstein} on Brownian motion and the first
statement of a Fluctuation-Response theory, through its celebrated
relation between diffusivity and mobility; the Onsager's works on the
reciprocity of transport coefficients and on the regression hypothesis
of large fluctuations~\cite{onsager}; and the works of Green and Kubo
on the general linear response formalism and the
Fluctuation-Dissipation relation (FDR) in its full dynamical
form~\cite{green,kubo}. All these works have resulted in the
construction of the so-called Non-Equilibrium
Thermodynamics~\cite{degroot}, that studies non-equilibrium states of
smoothly varying locally equilibrated regions, i.e.\ of systems in
local thermodynamic equilibrium.

The FDR links the mean response to a perturbation with a suitable
correlation function computed for the unperturbed system, and was
originally developed in the context of (equilibrium) statistical
mechanics of Hamiltonian systems. This fact was misunderstood by some
authors. For instance, it was claimed (with qualitative arguments)
that in fully developed turbulence (which is a non Hamiltonian and non
equilibrium system) there is no relation between fluctuations and
relaxation to the statistical steady state~\cite{Rose78}.  In reality,
a generalized FDR holds under rather general hypothesis, independently
from the Hamiltonian nature of the systems~\cite{F90, Falc95,
boffy02}.  If the system is mixing and the invariant measure is
``smooth'' enough, there exists a connection between the ``non
equilibrium'' properties (response to external perturbations) and the
``equilibrium'' properties (correlation functions computed according
to the equilibrium measure of the unperturbed system).  The validity
of the FDR does not depend on the deterministic or stochastic nature
of the system neither on the ``equilibrium'' or ``non equilibrium''
character of its statistical steady state.

In the last decade a series of important results have been obtained,
all related to a seemingly very general Fluctuation Relation (FR),
that constrains without any fitting parameter the pdf of the
fluctuations of entropy production or, more generally, of dissipated
energy. The main steps, in historical order, are: the proposal and the
observation of FR in numerical simulations, by Evans, Cohen and Morris
in 1993~\cite{evanscohenmorriss}, a derivation of the same relation in
transient states in 1994, by Evans and Searles~\cite{es}, and the
derivation of a theorem for asymptotic states by Gallavotti and Cohen
in 1995~\cite{gallavotticohen}.

If the chaotic hypothesis of the Gallavotti-Cohen theorem holds, the
deterministic dynamics can be mapped onto a mixing Markovian
stochastic process: this observation stimulated the derivation of a
stochastic version of the FR, first obtained by
Lebowitz and Spohn~\cite{lebowitz} and, more rigorously (and under more
restrictive hypothesis) by Maes~\cite{maes}. The case of Langevin
equations has been discussed by Kurchan~\cite{kurchan}.


In this paper we will work in the stochastic framework, where one can
directly speak of the probability, $\mathcal{P}_\tau(\Omega)$ say, of
observing a given trajectory segment $\Omega$ of temporal length
$\tau$. To begin with, we restrict our discussion to the simpler case
of discrete states and, therefore, discrete trajectories. Then, in the
presence of an appropriate form of reversibility, for each trajectory
$\Omega$ one can define the quantity
\begin{equation} \label{generic}
W'(\Omega)=\log \frac{\mathcal{P}_\tau(\Omega)}{\mathcal{P}_\tau(\overline{\Omega})},
\end{equation}
where $\overline{\Omega}$ is the reversal of $\Omega$~\footnote{In section II
the quantity denoted by $W'$ will be contrasted with a different quantity
denoted simply by $W$.}.
When detailed balance is satisfied one has $W'(\Omega) \equiv 0$. In the 
following we will refer to such a case as to ``equilibrium''. Equation
(\ref{generic}) implies
$\mathcal{P}_\tau(\Omega)=\exp(W'(\Omega))\mathcal{P}_\tau(\overline{\Omega})$
and, straightforwardly,
\begin{eqnarray}
{\textrm{prob}}(W'=X)=\sum_{\Omega:W'(\Omega)=X}
\mathcal{P}_\tau(\Omega)&=&\sum_{\Omega:W'(\Omega)=X}\exp(W'(\Omega))\mathcal{P}_\tau(\overline{\Omega})=\nonumber \\\exp(X)\sum_{\Omega:W'(\Omega)=-X}\mathcal{P}_\tau(\Omega)&=&\exp(X){\textrm{prob}}(W'=-X).
\end{eqnarray}
The symmetry ${\textrm{prob}}(W'=X)=\exp(X){\textrm{prob}}(W'=-X)$ is the
simplest form of a FR. This relation is the basis of a variety of FR's
discussed in the recent literature~\cite{frs}. 

In the treatment of Gallavotti and Cohen for deterministic systems, the time 
$\tau$ must be large, the role of $W'$ is played by the time-integrated phase 
space contraction rate and the following large deviation function of $W'$ is 
considered: 
\begin{equation}
\pi(x)=\lim_{\tau \to \infty}\frac{1}{\tau}
\log({\textrm{prob}}(W' \in ((x-\delta)\tau,(x+\delta)\tau) ~, \qquad 
\mbox{for any }~ \delta > 0 ~.
\end{equation}
The FR for $W'$ can then be written as 
\begin{equation}
 x - \delta \le \pi(x)-\pi(-x) \le x + \delta.
\end{equation}

The idea underlying definition~\eqref{generic} is to identify
heuristically the contribution of a single trajectory to the entropy
production, so that $W'$ physically represents an integrated flux
(e.g. heat flux) divided by a temperature (or a sum of fluxes divided
by the temperatures of the different reservoirs involved),
i.e. $W'=\beta E Y$ where $E$ is the applied non-conservative field,
$Y$ is its conjugated observable and $E \times Y$ is an energy. For
small Gaussian fluctuations, the FR  yields directly $\langle W'^2 \rangle_c
\equiv \langle W'^2 \rangle-\langle \oW\rangle^2 = 2\langle W'
\rangle$, because of the parabolic shape of
the large deviation function. Then the classical FDR emerges as a particular case near
equilibrium (i.e. near $\langle W' \rangle =0$). In fact, the previous
physical identification of $W'$ implies $\beta \langle Y^2 \rangle_c =
2 \langle Y \rangle/E$. Therefore, assuming that $\langle Y^2
\rangle_c=2Dt$ where $D$ is given by its equilibrium value (as it is true
to first order in $E$) and $\langle Y \rangle=mEt$ (linear response),
one obtains $m=\beta D$ which is the celebrated Einstein relation
between mobility $m$ and diffusivity $D$. In a similarly
straightforward fashion, the Onsager reciprocal relations can be
derived from the FR, recalling the time reversibility of the
evolution.


In deterministic dynamics, the Gallavotti-Cohen theorem is based on
the hypothesis that the system is Anosov, hence that its phase space
is bounded. In some systems (for example in presence of singular
potentials) this hypothesis is not well controlled, because
non-conservative forces let $W'$ fluctuate without bounds.  In the
stochastic framework, this problem can be cast into simple
mathematical terms. In particular, Ref.~\cite{lebowitz} considers a
Markov chain, in the stationary state, with probability of each
trajectory $\Omega$ given by
\begin{equation}
\mathcal{P}(\Omega)=\mu_{\sigma_0}\Pi_{i=0}^{\tau-1} K_{\sigma_i \sigma_{i+1}}
\end{equation}
where $\mu_{\sigma_0}$ is the invariant probability of the state
$\sigma$ and $K_{\sigma_i \sigma_{i+1}}$ is the transition probability
associated with the jump $\sigma_i \to \sigma_{i+1}$ between two possible 
states of the chain. Therefore, the functional $W'$ can be divided in two 
pieces:
\begin{equation}
W'=W+B
\end{equation}
where $B$ is $\log\mu_{\sigma_0}/\mu_{\sigma_\tau}$. In this paper we
will argue that the fluctuations of $B$, which has to be considered as
a ``border'' term, are of practical importance (i.e. in numerical and,
presumably, also in real experiments) in the measurements of $W$.  The
relevance of border terms for finite and infinite times has been
discussed recently in other contexts: Farago~\cite{farago} was the
first, to the best of our knowledge, to show with exact results the
difference of the large deviations of work done by the thermostat and
energy dissipated by the system, in stochastic models described by a
Langevin equation. Border terms had been dealt with also in
deterministic dynamics, in the paper~\cite{evansadv} by Evans and
Searles, where the "dissipation function" $f$, analogous to our $W'$, was
introduced as a physically relevant quantity. In a time dependent
Langevin equation, van Zon and Cohen~\cite{vanzon} and more recently
Baiesi and co-workers~\cite{maes_new} have studied the difference
between work done and heat transferred to the thermostat; the effect
of a border term making the difference between two heat fluxes in a
non-equilibrium Langevin model has been studied with exact analytical
results in a more recent paper by Visco~\cite{visco}. Simulations of
deterministic system~\cite{kovacs} have revealed the importance of
border terms also at finite times.
In Ref.~\cite{galla_last} the necessity of the inclusion of border
terms was recognized, and 
the example of a Bernoulli process has been discussed.  In the
Lebowitz-Spohn formulation, the problem of borders at finite times has
been treated in~\cite{puglisitracer}, considering simulations of a
tracer particle in a granular gas. The necessity of
including border terms has also been recognized in~\cite{seifert}.

The term $B$ can contribute to the large deviations of $W$, even if
all time-rescaled cumulants of its fluctutations vanish.  Van Zon and
Cohen~\cite{vanzon} first made a general observation: if
one has a time-extensive variable $Y(t)$ (an observable whose
cumulants all grow linearly in time~\footnote{We informally use the
term extensive, referring to the ``time-Gibbsian property'' (see for
example~\cite{maes}) of the measure of $\mathcal{P}(\Omega)$ and
therefore of $\textrm{prob}(W=X)$: this amounts to say that the
probability of $W$ obeys a large deviation scaling, being time the
``large parameter'', or in other words that it is characterized by an
exponential decay when the time length of the trajectory is linearly
increased.}) and a non extensive variable $X$ with unbounded
fluctuations whose probability has exponential tails, or slower, then
there is no guarantee that the extensive variable $Z(t)=X+Y(t)$ has
the same large deviations of $Y(t)$. Heuristically, this observation
may descend from the existence and finiteness of the large deviation
function for $X$,
\begin{equation}
\pi(x)=\lim_{t \to \infty}\frac{1}{t}\log 
\textrm{prob}(X \in ((x-\delta)t,(x+\delta)t) ~.
\end{equation}

On practical grounds, we will show that the ``memory'' of initial and
final conditions, i.e. the relevance of $B$, lasts much longer ($10^3$
times longer and more) than the typical time defined in terms of the
probability evolution of the Markov process toward the invariant
probability, making the FR for $W$ practically impossible to verify in
almost all our examples. Recent theoretical studies~\cite{maes_new}
predict that the form of the violation of the FR for W should have
given character in general, so that the resulting FR would take the
form of the heat FR of Van Zon and Cohen. These studies concern a
given class of systems and observables, hence it is interesting to see
which other systems verify the same predictions and which systems do
not. For instance, Ref.~\cite{harris} shows models in which the violation of
the FR takes on a different form. Therefore, we investigate also this
question.

In section II we will pose the problem in a general form. Our two
stochastic examples, a Markov chain with a large
number of states and a continuous time Markov process
exactly defined by the dynamics of a tracer in a granular
gas~\cite{puglisitracer}, will be discussed respectively in Section 
III and IV. Concluding remarks will be given in section V.

\section{Action functionals in Markov chains}

\subsection{Definitions and properties}

Consider a Markov chain with $N$ (possibily infinite) states with
invariant probability measure $\bmuu$ and transition rates $K_{ab}$ which denote
the conditional probability of going from state $a$ to state $b$, we
are interested in the fluctuations of the following two ``action
functionals'':
\begin{eqnarray} 
\label{eq:functionals}
W(\tau)&=&\sum_{i=0}^\tau \log\frac{K_{\sigma_i \sigma_{i+1}}}{K_{\sigma_{i+1}
   \sigma_i}}\\
\oW(\tau)&=&\sum_{i=0}^\tau \log\frac{K_{\sigma_i  \sigma_{i+1}}}{K_{\sigma_{i+1}  \sigma_i}}+\log\frac{\mu_{\sigma_0}}{\mu_{\sigma_{\tau+1}}}
\end{eqnarray}
where $\sigma_i \in \{1, 2, ... N\}$ is the state of the system at time $i$. Both
functionals have been defined by Lebowitz and Spohn, in~\cite{lebowitz}, who 
focused on the properties of $W$, neglecting the importance of the
difference $B=\oW-W$. For this reason, we call $W$ as the ``Lebowitz-Spohn 
functional'', while we refer to $\oW$ as to the ``corrected'' or ``adjusted'' 
functional. Both $W$ and $\oW$ associate a real number to any finite trajectory
(any realization of the Markov chain), and have the following properties:

\begin{itemize}
  
\item {\bf Zero at Equilibrium}: by equilibrium we mean a stationary
  state which verifies the detailed balance condition $\mu_a K_{ab}=\mu_b K_{ba}$, 
  being $\bmuu=\{\mu_i\}$ the invariant measure. Then, one has $W(t) \equiv 0$ 
  if   $K_{ab}=K_{ba}$, even during transient states represented by measures
  other than the microcanical one. However, $W(t) \ne 0$ even at equilibrium
  except for isolated (microcanonical) systems. Differently, 
  $\oW(t) \equiv 0$ at any equilibrium state, but not in transient states. 
  
\item {\bf Ergodicity}: for $t$ large enough, for almost all the
  trajectories $\lim_{s \to \infty} W(s)/s = \lim_{s \to \infty}
  \oW(s)/s=\langle W(t)/t \rangle=\langle \oW(t)/t \rangle$; here
  (assuming an ergodic and stationary system)  $\langle
  \rangle$ indicates an average over many independent segments from
  a single very long trajectory.
 
\item {\bf Entropy production}: Let $S(t)=-\sum_{i=1}^N \nu_i \log
  \nu_i$ be the entropy of the system at time $t$, where $\nu_i(t)$ is 
  the probability to be in the state $i$ at time $t$; then

\begin{equation}
S(t+1)-S(t)=R(t)-A(t)
\end{equation}

 where $R(t)$ is always non-negative, $A(t)$ is a linear function with
 respect to $\bnuu(t)=\{\nu_i(t)\}$, and $\langle W(t)
 \rangle=\langle \oW(t) \rangle \equiv \int_0^t \dd t'
 A(t')$. In~\cite{lebowitz} this has been shown for continuous time
 Markov processes, but the proof is valid also in the discrete time
 case (see for example~\cite{gaspard04}). This leads to consider
 $W(t)$ and $\oW(t)$ equivalent to the contribution of a single trajectory 
to the total entropy flux. In a stationary state
 $A(t)=R(t)$ and therefore the flux is equivalent to the
 production. 

\item {\bf Positivity of the average}: for large enough $t$: 1) at
  equilibrium (i.e. when there is detailed balance) $\langle W(t)
  \rangle=\langle \oW(t)\rangle=0$; 2) out of equilibrium those two
  averages are positive.

\end{itemize}

Let us now introduce our objects of study:

\begin{itemize}

\item {\bf $FR_W$}:
  $\pi(w)-\pi(-w)=w$ where $\pi(w)=\lim_{t\to \infty}\frac{1}{t}\log{f(t,t
    w)}$ and $f(t,x)$ is the probability  of finding
  $W(t)=x$ at time $t$. 
  Note that, in
  principle $\Pi(t,w)=\frac{1}{t}\log{f(t,t w)}\neq\pi(w)$ at any finite
  time; a derivation of this property has been obtained
  in~\cite{lebowitz}, while a rigorous proof with more restrictive hypothesis
  is in~\cite{maes}; the discussion for the case of a Langevin equation is
  in~\cite{kurchan}.
  
\item {\bf $FR_{\oW}$}:
  $\oPi(t,w)-\oPi(t,-w)=w$ where
  $\oPi(t,w)=\frac{1}{t}\log{\of(t,t w)}$ and
  $\of(t,x)$ is the probability density function of finding
  $\oW(t)=x$ at time $t$.

\end{itemize}


\subsection{Fluctuations of W}

We first define an extended probability vector $\bp(t,W)$ where each
component $p_i(t,W)$ is the probability of finding
the system at time $t$ in the state $i$ with the value $W$ for the
Lebowitz and Spohn functional. This means that $\sum_i
p_i(t,W)=f(t,W)$ is the probability density function at time $t$ for
the functional $W$, while $\int \dd W p_i(t,W)=\nu_i(t)$ is the
probability of finding the system in state $i$ at time t and $\int \dd
W \sum_i p_i(t,W)=1$.  We will use the simple notation $\mu_i$ to
denote the invariant measure, i.e. $\mu_i= \lim_{t\to \infty}
\nu_i(t)$.  The evolution of $\bp(t,W)$ is given by the following
equation:

\begin{equation}
p_i(t+1,W)=\sum_j K_{ji}p_j(t,W-\Delta w_{ji})
\end{equation}
where $K$ is the previously defined transition matrix and $\Delta
w_{ij}$ is the variation of $W$ due to a jump from the state $i$ to
the state $j$. This reads $\Delta w_{ij}=\ln \frac{K_{ij}}{K_{ji}}$
for the original Lebowitz-Spohn functional.

Then define the function $\bpt(t,\lambda)=\int \dd W
\exp(-\lambda W) \bp(t,W)$ and obtain, for its evolution
\begin{equation}
\pti_i(t+1,\lambda)=\sum_j K_{ji}\pti_j(t,\lambda)e^{-\lambda \Delta w_{ji}}.
\end{equation}
In a compact form we can write
\begin{equation}
\bpt(t+1,\lambda)=A(\lambda) \bpt(t,\lambda)
\end{equation}
$A(\lambda)$ being the matrix defined by
\begin{equation}
A_{ij}(\lambda)=K_{ji}^{1-\lambda}K_{ij}^\lambda,
\end{equation}
and, therefore,
\begin{equation}
\bpt(t,\lambda)=A(\lambda)^t\bpt(0,\lambda)
\end{equation}
with $\pti_i(0,\lambda)=\int \dd W \exp(-\lambda W) \nu_i(0)\delta(W)=\nu_i(0)$.

Note that the $ij$-term of the evolution matrix
$\ft_{ij}(t,\lambda)=[A(\lambda)^t]_{ij}$ is a function of $\lambda$
that represents the characteristic function of the distribution of $W$
at time $t$ constrained by the conditions $\sigma_1=j$ and
$\sigma_t=i$ (i.e. the system is in state $j$ at time $0$ and in state
$i$ at time $t$). In turn, the characteristic function of the
distribution of $W$ is obtained summing over all the states:
\begin{equation}
\ft(t,\lambda)=\sum_i \pti_i(t,\lambda)=\sum_i 
\sum_j [A(\lambda)^t]_{ij}\nu_j(0)=\sum_j \nu_j(0) \sum_i [A(\lambda)^t]_{ij}.
\end{equation}

In principle (under suitable analiticity assumptions) one recovers the
distribution of $W$ at time $t$ by inverting the transform:
\begin{equation}
f(t,W)=\int_{-i \infty}^{+i \infty} \dd\lambda \exp(\lambda W) \ft(t,\lambda).
\end{equation}
The knowledge of the characteristic function suffices to generate the 
moments or the cumulants of $W$: 

\begin{eqnarray}
 \label{cumulants}
\langle W_t^n \rangle&=&(-1)^n\left . \frac{d^n}{d\lambda^n}
\ft(t,\lambda) \right|_{\lambda=0}\\
\langle W_t^n \rangle_c&=&(-1)^n\left .  \frac{d^n}{d\lambda^n}
\log \ft(t,\lambda) \right|_{\lambda=0}.
\end{eqnarray}

\subsubsection{The large time behavior and the $FR_W$}

At large times, the evolution operator $A(\lambda)^t$ is dominated by
the largest eigenvalue $y_1(\lambda)$ of $A(\lambda)$. Defining
$y_1(\lambda)=\exp(\zeta(\lambda))$ it follows that
\begin{eqnarray}
\label{largetimes}
\bpt(t,\lambda) &\sim& \exp(\zeta(\lambda) t) \left[\sum_j x^{(1)}_j(\lambda) \nu_j(0)\right]
\mathbf{x}^{(1)}(\lambda)\\
\tilde{f}(t,\lambda)=\sum_i \pti_i(t,\lambda) &\sim& \exp(\zeta(\lambda)t) \left[\sum_j x_j^{(1)}(\lambda) \nu_j(0)\right] \sum_i x^{(1)}_i(\lambda) ,
\end{eqnarray}
where $\mathbf{x}^{(1)}(\lambda)$ is the eigenvector of $A(\lambda)$
associated to the largest eigenvalue $y_1(\lambda)$.  One expects,
from the above large times behavior, an analogous large time behavior
for the density distribution of $W$, i.e.

\begin{eqnarray} \label{saddle}
f(t,W)=\int_{-i \infty}^{+i \infty} \dd\lambda \exp(\lambda W) \sum_j \nu_j(0) \sum_i [A(\lambda)^t]_{ij} \nonumber \\
\sim \exp\left[t \max_\lambda\left(\lambda \frac{W}{t}+\zeta(\lambda)\right)\right]=\exp[t \pi(W/t)]
\end{eqnarray}
where we have defined $\pi(w)=\lim_{t \to \infty} \frac{1}{t}\log
f(t,wt)$, the large deviation function associated with $f(t,W)$, which
(under the validity of the last chain of equalities) is obtained as
a Legendre transform of $\zeta(\lambda)$, i.e. $\pi(w)=\lambda^*
w+\zeta(\lambda^*)$ with
$\frac{d}{d\lambda}\zeta(\lambda)|_{\lambda=\lambda^*}=-w.$

The $FR_W$ symmetry appears at this
stage. In fact, it is evident that $A(\lambda)=A^T(1-\lambda)$ being
$A^T$ the transposed of $A$. This implies that
$\zeta(\lambda)=\zeta(1-\lambda)$ which suffices to get
$\pi(w)=\pi(-w)+w$, i.e. $f(t,W)=f(t,-W)\exp(W)$ at large times.

Clearly the validity of the
expansion in~\eqref{saddle} is crucial for the validity of the $FR_W$
symmetry, but it is not guaranteed when the
integrand presents non-analiticities in the $\lambda$ complex
plane. Such a catastrophe can happen, for example, when the number of
states $N$ becomes infinite and the initial (final) probability
$\mu_{j(i)}(0)$ has some unbounded form (see below). The physical
meaning of such a catastrophe is that the large fluctuations in the 
initial and final state cannot be neglected, because they contribute 
to the tails (i.e. the large deviations) of $f(t,W)$ at any time and, 
therefore, they prevent the $FR_W$ symmetry.

The problem, now, is to find a condition sufficient to trigger the
analyticity breaking, in terms of initial/final state distribution. 
Note that the breaking of analiticity of
$\ft(t,\lambda)$ is associated with large fluctuations of invariant
probabilities, but here ``large'' has a very generic meaning and does
not require the presence of any power law tail. Apparently (from many
studies including numerical simulations~\cite{puglisitracer}, analytical 
calculations~\cite{farago,visco} and heuristic arguments, cf.\ for
example~\cite{vanzon}), the initial and final configurations can be a 
problem when the distribution of the boundary term $B=\ln \mu_{\sigma(0)}
-\ln \mu_{\sigma(t)}$ has exponential, or higher, tails in the stationary 
state. But the problem may occur {\em even if $\mu_i$ has Gaussian tails}!

\subsubsection{The second functional and the finite time symmetry relation}

Let us see what happens to the ``adjusted'' functional $\oW$. Here
the prime will always denote quantities which involve this second
functional.  For example we have the extended measure vector
$\obp(t,\oW)$ and the probability density function at time
$t$, $\of(t,\oW)$, the characteristic function $\oft(t,\lambda)$, etc.
The increment for the functional at each jump reads $\Delta
w'_{ij}=\Delta w_{ij}\log\left(\frac{\mu_i}{\mu_j}\right)$,
recalling that $\mu_i$ is the invariant measure for the state $i$. The evolution matrix therefore reads
\begin{equation}
\oA_{ij}(\lambda)=K_{ji}^{1-\lambda}K_{ij}^\lambda \mu_j^{-\lambda} \mu_i^\lambda,
\end{equation}
and the evolution equation reads
\begin{equation}
\obpt(t+1,\lambda)=\oA(\lambda) \obpt(t,\lambda).
\end{equation}
The fundamental difference between $\oW$ and the original $W$ function appears
now. Since $\obpt(0,\lambda)=\bnuu(0)$, if one takes
$\bnuu(0)\equiv\bmuu$ (the system is in the stationary regime from the beginning),
it happens that
\begin{equation}
\oft(1,\lambda)=\sum_i \opt_i(1,\lambda)=\sum_i \sum_j \oA_{ij}(\lambda) \mu_j=\sum_i \sum_j K_{ji}^{1-\lambda}K_{ij}^\lambda \mu_j^{1-\lambda} \mu_i^\lambda,
\end{equation}
i.e. $\oft(1,\lambda)=\oft(1,1-\lambda)$. By recursivity, one realizes
that this is the case for all times $t$, i.e.\
\begin{equation}
\oft(t,\lambda)=\oft(t,1-\lambda),
\end{equation}
in general, which leads immediately to a finite-time symmetry relation
\begin{equation}
\of(t,\oW)=\of(t,-\oW)\exp(\oW),
\end{equation}
valid for any $t$.

The second functional contains a term that absorbs all
the effects of the fluctuations of the steady state measure, leading to a
conservation of the symmetry $\lambda \to 1-\lambda$ all along the
evolution. Such a conservation prevents bad surprises at large
times also in presence of large fluctuations of the initial measure.

\subsubsection{Characteristic times}

\label{times}

When discussing the asymptotic validity of a fluctuation relation in a
finite time simulation an immediate question arises: what are the
characteristic times of the evolution and how large with respect to
them is the chosen time $\tau$? The correct answer, here, is that one
is really probing the asymptotic behavior, i.e. really measuring the
large deviations of functionals $W$ and $\oW$, if and only if the
fluctuations satisfy the following time scaling:
\begin{equation} \label{scaling}
f(t,W) \sim \exp[t \pi(W/t)].
\end{equation}
Actually, this is equivalent to observe stationary values for the
time-dependent cumulants rescaled with the time $c_n(t)=\langle W(t)^n
\rangle_c/t$ and $\ovc_n(t)=\langle \oW(t)^n \rangle_c/t$. Verifying
the stationarity of the rescaled cumulants sometimes proves to be
advantageous with respect to the scaling~\eqref{scaling}: a cumulant
is in fact a simple number containing information on $f(t,W)$ in an
integrated form. On the other hand, the stationarity of a cumulant is
a necessary but not sufficient condition for the validity of the
scaling~\eqref{scaling} (all cumulants must be stationary) and its
measure may eventually be very noisy.

The large time dominance of the maximum eigenvalue allows 
the direct calculation of the asymptotic cumulants of $W$ and $\oW$,
i.e.
\begin{eqnarray}
\label{asymptotics}
c_n(\infty)=\lim_{t \to \infty}\langle W^n \rangle_c/t &=&(-1)^n\left . \frac{d^n}{d\lambda^n}
\zeta(\lambda) \right|_{\lambda=0}\\
\ovc_n(\infty)=\lim_{t \to \infty}\langle (\oW)^n \rangle_c/t &=&(-1)^n\left . \frac{d^n}{d\lambda^n}
\ozeta(\lambda) \right|_{\lambda=0}.
\end{eqnarray}
These formulae are valid provided that $\zeta(\lambda)$ and
$\ozeta(\lambda)$ are analytic in zero. Numerical observations in the
following example suggest a coincidence between $c_n(\infty)$ and
$\ovc_n(\infty)$. This is consistent with direct calculations
of $\zeta(\lambda)$ and $\ozeta(\lambda)$, which appear to
coincide, in Markov chains with a small number of states. We have 
not been able to address analytically the case of a
Markov chain with an infinite number of states. The other evidence
which will appear from the simulations (already noticed in the
granular tracer model~\cite{puglisitracer}) is that the cumulants of
$W$ are much slower than those of $\oW$ to converge to a
stationary value. 

The time of convergence to the perfect large deviation
scaling~\eqref{scaling} is determined in a complex manner by the other
eigenvalues (from the second one) of the operator $A(\lambda)$ (or
$\oA(\lambda)$). It will be interesting to compare this convergence
time (estimated for example from the convergence time of the rescaled
cumulants) with the characteristic time of the Markov chain itself
$\tau_M$. We will use, for $\tau_M$, the time obtained with the
following standard recipe. An initial vector $\bnuu(0)$ evolves under
the action of the transpose of the transition matrix:
\begin{equation}
\bnuu(t)=(K^T)^t\bnuu(0).
\end{equation}
For a generic $\bnuu(0)$ one has $\bnuu(t)=\bmuu+0({\textrm e}^{-t/\tau_M})$
where the characteristic time $\tau_M$ is given by the second
eigenvalue $\alpha_2$ of $K^T$, i.e. the closest one to the unitary
circle:
\begin{equation}
\tau_M=\frac{1}{|\ln(\alpha_2)|}.
\end{equation}
One can easily compute $\tau_M$ from the area $a(t)$ of the
parallelogram identified by $\bmuu$ and $\bnuu(t)$:
\begin{equation}
a(t) \sim \exp(-t/\tau_M).
\end{equation}

\section{A Markov chain}

We discuss a simple Markov chain which does not
correspond to any particular physical system, but turns out to be useful
in illustrating the relevance of the boundary term $B$
in the fluctuations of the two action functionals. Such a model has been
inspired by a work by Gaspard and Wang~\cite{gaspard}. 

\subsection{The model}

We consider a Markov chain with $N+2$ states, labelled $A$, $B$ and
$C_i$ with $i \in \{1, 2, ... N\}$.  The transition matrix $K$ is given by

\begin{equation} \left(
\begin{array}{llllll}
p_{AA}&   p_{AB}&  (1-p_{AA}-p_{AB})k_1& (1-p_{AA}-p_{AB})k_2&
\ldots& (1-p_{AA}-p_{AB})k_N\\
p_{BA}&    0&     (1-p_{BA})k_1& (1-p_{BA})k_2&   \ldots& (1-p_{BA})k_N\\
p_{CA}&       1-p_{CA}&     0& 0&              \ldots& 0\\
p_{CA}&       1-p_{CA}&     0& 0&              \ldots& 0\\
\vdots& \vdots&  \\
\end{array} \right)
\end{equation}
with $\sum_{i=1}^N k_i=1$, under the constraint that for every 
jump with nonzero transition probability, the reversed jump is 
also possible. The invariant probability $\boldsymbol{\mu}$ for 
$N=1$ is given by
\begin{eqnarray}
\mu_A&=&\frac{p_{BA}+p_{CA}-p_{BA}p_{CA}}{\mathcal{N}}\\
\mu_B&=&\frac{1-p_{AA}+p_{CA}(p_{AB}+p_{AA}-1)}{\mathcal{N}}\\
\mu_C&=&\frac{1-p_{AA}-p_{AB}p_{BA}}{\mathcal{N}}
\end{eqnarray}
with $\mathcal{N}=2+p_{AA}(-2+p_{CA})+p_{AB}p_{CA}-p_{BA}(-1 + p_{AB} + p_{CA})$.

It can be easily seen that the invariant measure for $N>1$ is
the same as in the case $N=1$ with a decomposition of the measure of state
$C_1$ into the measures of $C_i$ proportional to the values $k_i$:

\begin{equation}
\mu_{C_i}=\mu_C k_i
\end{equation}

\subsection{Numerical results}

\subsubsection{Fluctuations of the functionals}

\begin{figure}[htbp]
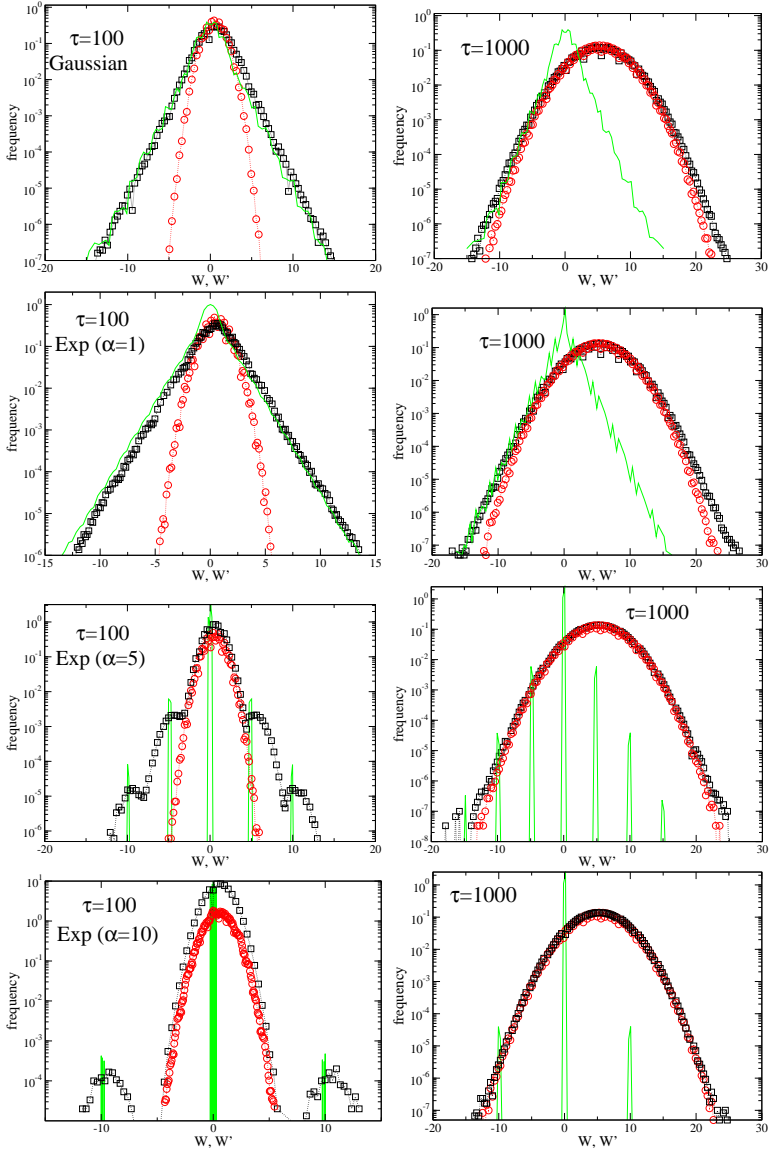

  \includegraphics[width=5cm,clip=true]{pdf_gauss_t100.eps}
  \includegraphics[width=5cm,clip=true]{pdf_gauss_t1000.eps}\\
  \includegraphics[width=5cm,clip=true]{pdf_exp1_t100.eps}
  \includegraphics[width=5cm,clip=true]{pdf_exp1_t1000.eps}\\
  \includegraphics[width=5cm,clip=true]{pdf_exp5_t100.eps}
  \includegraphics[width=5cm,clip=true]{pdf_exp5_t1000.eps}\\
  \includegraphics[width=5cm,clip=true]{pdf_exp10_t100.eps}
  \includegraphics[width=5cm,clip=true]{pdf_exp10_t1000.eps}
\caption{{\bf Color online.} Probabilities of observing $W$ (black
squares), $\oW$ (red circles) and $B=\oW-W$ (green lines), for the
Markov chain. Each line is composed of two graphs and shows the
results for a particular choice of the parameters: the left column is
at time $\tau=100$ and the right column at time $\tau=1000$. In all
the simulations we have used $N=50$. In all frames we have used
$p_{AA}=0.2$, $p_{AB}=0.3$, $p_{BA}=0.3$ and $p_{CA}=0.5$: the first
line corresponds to Gaussian tails of the invariant measure, because
$k_i \propto exp(-i^2/10)$, while the last three lines correspond
to exponential choices $k_i \propto exp(-\alpha i)$ with the value of
$\alpha$ given in the plot. In all cases $\langle W(\tau)
\rangle=\langle \oW(\tau) \rangle=0.0053\tau$, i.e. $0.53$ for
$\tau=100$ and $5.3$ for $\tau=1000$.  \label{fig:map_pdf}}.
\end{figure}

\begin{figure}[htbp]
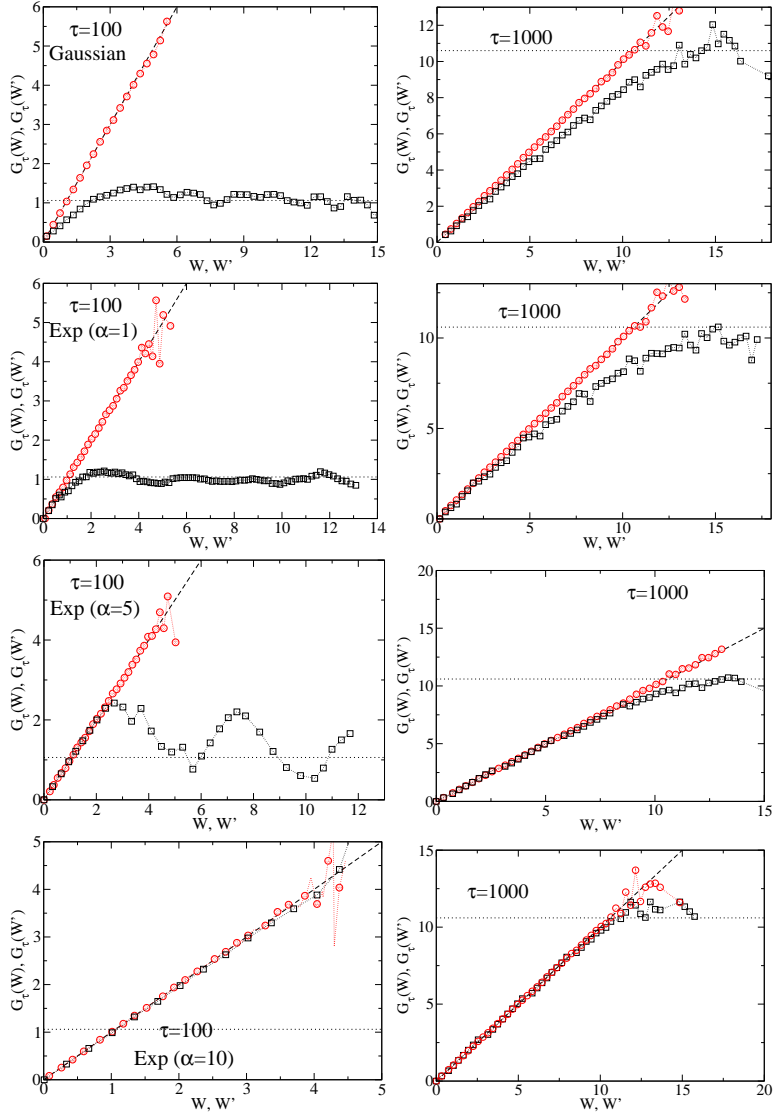

  \includegraphics[width=5cm,clip=true]{plot_gauss_t100.eps}
  \includegraphics[width=5cm,clip=true]{plot_gauss_t1000.eps}\\
  \includegraphics[width=5cm,clip=true]{plot_exp1_t100.eps}
  \includegraphics[width=5cm,clip=true]{plot_exp1_t1000.eps}\\
  \includegraphics[width=5cm,clip=true]{plot_exp5_t100.eps}
  \includegraphics[width=5cm,clip=true]{plot_exp5_t1000.eps}\\
  \includegraphics[width=5cm,clip=true]{plot_exp10_t100.eps}
  \includegraphics[width=5cm,clip=true]{plot_exp10_t1000.eps}
\caption{{\bf Color online.} $G_\tau(W)$ vs. $W$ (black squares) and
  $G_\tau(\oW)$ vs. $\oW$ (red circles) for the Markov chain. Each
  line is composed of two graphs and shows the results for a
  particular choice of the parameters: the left column is at time
  $\tau=100$ and the right column at time $\tau=1000$. The dashed line
  has slope $1$. In all the simulations we have used $N=50$. In all
  frames we have used $p_{AA}=0.2$, $p_{AB}=0.3$, $p_{BA}=0.3$ and
  $p_{CA}=0.5$: the first line corresponds to Gaussian tails of the
  invariant measure, because $k_i \propto exp(-i^2/10)$, while the
  last three lines correspond to exponential choices $k_i \propto
  exp(-\alpha i)$ with the value of $\alpha$ given in the plot. In all
  cases $\langle W(\tau) \rangle=\langle \oW(\tau)
  \rangle=0.0053\tau$, i.e. $0.53$ for $\tau=100$ and $5.3$ for
  $\tau=1000$. The dotted horizontal line marks the van Zon and Cohen
  prediction for the FR violation, $G_\tau(W)=2\langle W \rangle$ for large $W$~\cite{vanzon}.
  \label{fig:map}}.
\end{figure}

We measure the functionals $W(\tau)$ and $\oW(\tau)$  along
independent non-overlapping segments of time-length $\tau$ extracted from a
unique trajectory after the stationary regime has been achieved.
In particular we probe the validity of a relation like
\begin{equation}
  G_\tau(X)=\log F(\tau,X)-\log F(\tau,-X)=X \label{as_es}
\end{equation}
where $F(t,X)$ is the probability density function of finding one of
the two functionals $W$ or $\oW$ after a time $t$ equal to $X$
(i.e. $F(t,X)$ corresponds to $f(t,W)$ or to $f'(t,\oW)$ depending on
the cases). All the results are shown in figures~\ref{fig:map_pdf}
and~\ref{fig:map}, containing the graph of $G_\tau$ vs $x$ and the
pdfs of $W$ and $\oW$, with different choices of the transition
probabilities and of the time $\tau$. We separately discuss three main
cases of interest.

1) Whenever $A$ is disconnected from all $C_i$'s, i.e. when
$p_{AB}=1-p_{AA}$ and $p_{CA}=0$, detailed balance is satisfied. The
numerical results confirm what is expected: in the detailed balance
(equilibrium) case, $\oW(\tau)$ is identically zero and does not
fluctuate (its pdf is a delta in zero) and therefore $W(\tau)$
coincides with the opposite of the boundary term
$B=\log\frac{\mu_{\sigma_1}}{\mu_{\sigma_\tau}}$: they both have
symmetric fluctuations around zero with exponential tails, and      
$G_\tau \equiv 0$.

2) Lines 1-3 of figure~\ref{fig:map}. Whenever $A$ is connected to all
   $C_i$'s, i.e. when $p_{AB}<1-p_{AA}$ and $p_{CA}>0$, detailed
   balance is violated. In this case both $W(\tau)$ and $\oW(\tau)$
   fluctuate around a nonzero (positive) value which, for $\tau$ large
   enough, is the same for the two functions. We have chosen $N=50$, $p_{AA}=0.2$,
   $p_{AB}=0.3$, $p_{BA}=0.3$, $p_{CA}=0.5$, $\tau=100$.  In the first
   line of graphs of figure~\ref{fig:map}, we have $k_i \propto
   exp(-i^2/10)$, while in all the other frames $k_i \propto
   exp(-\alpha i)$ with different $\alpha$. The choice of the
   transition rates $k_i$ from state $B$ to state $C_i$ determines the
   invariant measure of states $C_i$, which is still proportional to
   $k_i$. Note however that, even in the Gaussian case, the
   fluctuations of the boundary term
   $B=\log\frac{\mu_{\sigma_1}}{\mu_{\sigma_\tau}}$ have a pdf with
   exponential tails, as confirmed by the numerical observation. The
   other relevant observation, here, is that the pdf of $\oW$ is
   almost perfectly Gaussian. This is not true for the pdf of $W$.  At
   small times ($\tau=100$) the pdf of $W$ resembles the pdf of $B$,
   evidencing that at this time $W$ is completely dominated
   by $B$. At larger times the pdf of $W$ and $B$ start to deviate, in
   particular that of $W$ tends to become equal to that of $\oW$,
   still with evidently different tails of exponential form. The tails
   of the pdf of $\oW$ seem to be always dominated by the fluctuations
   of $B$. Note that (in the third and fourth line of the figure) when
   the invariant measure is peaked on the first states (i.e. when
   $\alpha$ is high) the fluctuations of $B$ have the form of a sum of
   peaks and the pdf of $W=-B+\oW$ has the form of a sum of Gaussians
   centered on those peaks.

The symmetry relation for the functional $W$ is not verified in any of
these simulations, both at small and large times. The asymmetry
measure $G_\tau(W)$ has a slope near $1$ only for small values of $W
\ll \langle W \rangle$, then deviates and saturates to a constant
value in good agreement with the value $2\langle W \rangle$ predicted
by van Zon and Cohen~\cite{vanzon}, also in agreement with
ref.~\cite{galla_last}. Note that the observations at large times are
perfectly compatible with those at small times, i.e. if both ordinates
and abscisass are divided by $\tau$ the curves are similar (in the
first two cases they collapse very well). Therefore the ``reduction''
of the violation at large times is only apparent. On the other hand
the symmetry relation for $\oW$ is {\em always} satisfied, at all
times and for all the choices of the parameters.

3) Fourth line of figure~\ref{fig:map}. In this simulation the
   invariant measure on states $C_i$ is still an exponential, but with
   a higher slope. In this case our numerical results show that the
   fluctuations of $W$ are much closer to those of $\oW$ already at
   small times. This is reflected on the good agreement with the
   Fluctuation Relation of both functionals $W$ and $\oW$. Note that
   one can still be doubtful about this verification, because of the
   limited range of values of $W$ available, which can possibly hide a
   failure at larger values. The distribution of the boundary
   term in fact still has exponential tails (invisible at our
   resolution) and again it can be argued that these tails (being
   Gaussian those of $\oW$) will dominate at very large values.

\subsubsection{Characteristic times: scaling and cumulants}

Following the recipe given in section~\ref{times}, we have calculated
the characteristic time $\tau_M$ associated with the approach toward the
invariant probability of the Markov chain. Our general observation is
that its dependence upon the choice of the particular form of the
transition probabilities $k_i$, which couple the state $B$ with the
$N$ states $C_i$, is negligible. In particular, for all the choice of
parameters corresponding to the cases shown in figure~\ref{fig:map}, we
have measured $\tau_M \approx 2.1$, while in the equilibrium case
$p_{AA}=0.2$, $p_{AB}=0.8$, $p_{BA}=0.3$ and $p_{CA}=0$, we have
measured $\tau_M \approx 18$.

\begin{figure}[hbtp]
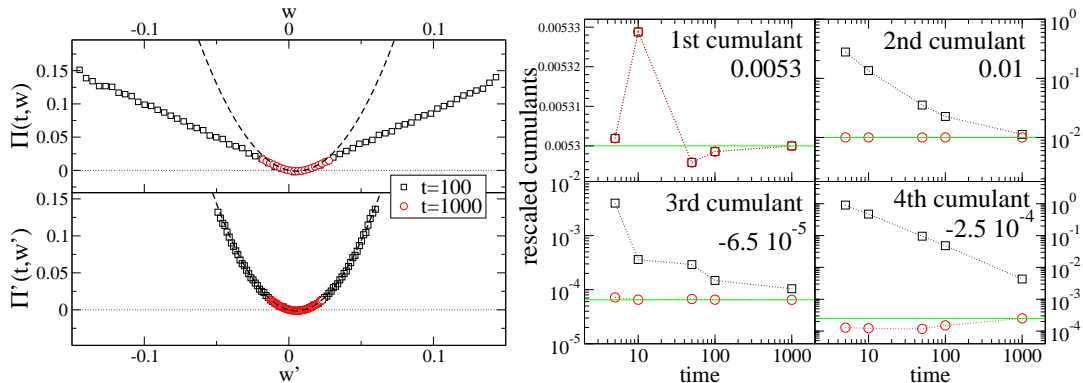

\includegraphics[height=5cm,clip=true]{scaling.eps}
\includegraphics[height=5cm,clip=true]{cums.eps}
\caption{{\bf Color online}. Left: the large deviation function rate
at finite times
$\Pi(\tau,x)-\Pi(\tau,0)=-\frac{1}{\tau}\log\frac{F(\tau,x)}{F(\tau,0)}$
for the fluctuations of $W/\tau$ (top frame) and of $\oW/\tau$ (bottom
frame). The dashed line shows the quadratic function associated with
the Gaussian of same mean and same asymptotic variance, while the
dotted line marks $0$. Right: the first four time-rescaled cumulants
of $W$ (squares) and $\oW$ (circles) as a function of time. The solid
green line represents the absolute value of the cumulant of $\oW/\tau$
at $\tau=1000$: this value is also indicated in each frame. The
parameters are the same as in figure~\ref{fig:map_pdf}, second line
(i.e. $k_i \propto \exp -i$).\label{fig:cum}}
\end{figure}

In figure~\ref{fig:cum}-left we plot the large deviation rate function
at finite times $\Pi(t,x)$ for the two functionals $W$ and $\oW$, in a
non-equilibrium case with a choice of $k_i$ such that the fluctuations
of the invariant measure are large, i.e. $k_i \propto \exp(-i)$ (the
$\alpha=1$ case discussed in the previous section). The range of
available values drops very rapidly as $t$ increases, and, most
importantly, the negative range tends rapidly to disappear. On the
basis of these results one can conclude that the
scaling~\eqref{scaling} is already reached at $t=100$. This is true
for both functionals, $W$ and $\oW$. The two large deviation rate
functions appear different for $W$ and $\oW$, they seem to have
linear and quadratic tails respectively.

In figure~\ref{fig:cum}-right we show the values of the first four
cumulants (divided by the time, i.e. $c_n(t)$ and $c_n'(t)$ as defined
previously) of $W(t)$ and $\oW(t)$, as a function of time $t$. The
absolute values have been taken, in order to use a logarithmic plot
and better appreciate the different orders of magnitude. In each frame
of the figure the value of $c_n'(1000)$ is given. One first observes
that the cumulants of $\oW$ become time-independent almost
immediately. The values of the third and fourth cumulant are
consistent with the previous observation that the fluctuations of
$\oW$ are almost Gaussian, and its large deviation rate function is
quadratic. The remarkable fact is that the cumulants of $W$ are much
slower to converge to a stationary value: as a matter of fact we
cannot conclude, in our case, that this convergence has been actually
reached apart from the first cumulant. The measure of the cumulants
points out the weakness of a naked-eye verification of the large
deviation scaling as the one obtained in figure~\ref{fig:cum}-left.

\section{The granular tracer}

Recently, both in experiments~\cite{feitosa} and in
simulations~\cite{aumaitre}, the validity of the $FR_W$ has been
studied in granular gases. A granular gas consists of inelastic
particles, typically one has a box containing macroscopic grains (like sand)
strongly vibrated in order to obtain a dilute and fluidized stationary
regime, which is far from equilibrium because of non-conservative
forces acting on and among the grains. In a series of other papers~\cite{us}
it has been shown that the prototypical model of granular gas, the gas
of inelastic hard spheres, cannot be used without some care as a
benchmark for the validity of the fluctuation relation: each single
collision violates time reversibility, so that for any trajectory in
the phase space of the whole system, its reversed trajectory {\em does
not exist}. This implies that the large deviations of entropic
functionals are defined only for positive valuess, and that the fluctuation relation cannot
be verified. However, focussing on the Markovian dynamics of a single tracer
particle immersed in a gas, one may circumvent the problem of strong
irreversibility and probe the validity of the FR in granular
gases~\cite{puglisitracer}.

\subsection{The model}

We consider the dynamics of a tracer granular particle in a
homogeneous and dilute gas of grains which is driven by an unspecified
energy source. The tracer particle experiences consecutive collisions
with particles of the gas coming from the same ``population'',
independently of the position and time of collision. The gas is
characterized by its velocity probability density function
$P(\mathbf{v})$, which is well known to be mildly non-Gaussian and
well reproduced by a first Sonine correction~\cite{kicks}. The
dilution of the gas guarantees that the transient as well as the
stationary regimes of the velocity probability density function
$P_*(\mathbf{v})$ of the tracer is governed by a Linear Boltzmann
equation.

The inelastic collisions with the gas particles, which determine the
instantaneous changes of the velocity of the tracer, are described by
the simplest and most used inelastic collision rule:
\begin{equation}
\mathbf{v}'=\mathbf{v}-m\frac{1+\alpha}{m+M}[(\mathbf{v}-
\mathbf{V})\cdot \bo] \bo
\end{equation}
where $\mathbf{v}$ and $\mathbf{v}'$ are the velocities of the tracer
before and after the collision respectively, $\mathbf{V}$ is the
velocity of the gas particle, $m$ and $M$ are the masses of the tracer
and of the gas particle respectively, $\bo$ is the unitary vector
joining the centers of the two particles, and $\alpha$ is the
restitution coefficient that takes values in $[0,1]$ ($1$ is the
elastic case). For simplicity, in the rest of the discussion we will
consider 2-dimensional systems with $m=M$ (see~\cite{puglisitracer}
for a more general discussion).

The analysis of the Linear Boltzmann equation leads to the
following Master equation for the evolution of the velocity
probability density function of the tracer:

\begin{equation} \label{eq:markov}
\frac{\dd P_*(\mathbf{v},t)}{\dd t}=\int \dd \mathbf{v}_1 \left[ P_*(\mathbf{v}_1) K(\mathbf{v}_1, \mathbf{v}) -  P_*(\mathbf{v}) K(\mathbf{v}, \mathbf{v}_1) \right].
\end{equation}
where $P_*(v)$ is the velocity pdf of the test particle.  The
transition rate $K(\mathbf{v}, \mathbf{v}')$ (see~\cite{puglisitracer}
for a precise definition) of jumping from $\mathbf{v}$ to
$\mathbf{v}'$ is given by the following formula:
\begin{equation}
\label{eq:K}
K(\bv,\bv')=\left(\frac{2}{1+\alpha}\right)^2
\int \dd v_{2 \tau} P[\mathbf{v}_2(\mathbf{v},\mathbf{v}',v_{2 \tau})],
\end{equation}
where $\Delta \bv=\bv' - \bv$ denotes the change of velocity of the
test particle after a collision and $P(\mathbf{v})$ is the velocity
pdf of the gas. In Eq.~\eqref{eq:K} it has been assumed that the mean free
path equals $1$, which can be always obtained rescaling the
time. This means that the elastic mean free time is
$\tau_c^{el}=1/2\sqrt{\pi}$ (it is larger in inelastic cases). The
vectorial function $\bv_2$ is defined as
\begin{equation}
\mathbf{v}_2(\mathbf{v},\mathbf{v}', v_{2\tau})=v_{2\sigma}
(\mathbf{v},\mathbf{v}')\bsi(\mathbf{v},\mathbf{v}')+
 v_{2\tau} \bt  \label{eq:Kb},
\end{equation}
where $\bsi(\mathbf{v},\mathbf{v}')$ is the unitary vector parallel to
$\Delta \bv$ while $\bt$ is the unitary vector perpendicular to
it. Finally, to fully determine the transition rate (\ref{eq:K}), the
expression of $v_{2 \sigma}$ is needed:
\begin{equation}
v_{2 \sigma}(\bv , \bv') = \frac{2}{1+ \alpha} |\Delta \bv| 
+ \bv \cdot \bsi \,\,\,.
\end{equation}

For the purpose of measuring the fluctuations of the functionals
defined by Eq.~\eqref{eq:functionals}, Eq.~\eqref{eq:K} yields:
\begin{equation}
\frac{K(\mathbf{v},\mathbf{v}')}{K(\mathbf{v}',\mathbf{v})}=
\frac{\int \dd v_{2\tau}P[\mathbf{v}_2(\bv, \bv',v_{2\tau})]}{\int \dd v_{2\tau}P[\mathbf{v}_2(\bv', \bv,v_{2\tau})]}
\equiv  \frac{P[v_{2 \sigma} (\bv, \bv')]}{P[v_{2 \sigma} (\bv', \bv)]}\,\,\,.
\end{equation}
Then we consider two possible choices for the velocity pdf of the gas,
a Gaussian and a non-Gaussian case expressed in the form of a first
Sonine correction to the Gaussian~\cite{sonine}, which amounts to
$P(\mathbf{v})=\frac{1}{(2\pi T)}\exp{ \left( -
\frac{v^2}{2T}\right)}(1+a_2S_2^2(v^2/2T))$. The first Sonine
correction enters by means of the small parameter $a_2$ (which is $0$
in the Gaussian case) and the Sonine Polynomial
$S_2^{d}(x)=\frac{1}{2} x^2- \frac{d+2}{2} x+\frac{d(d+2)}{8}$, which
depends upon the choice of physical parameters, e.g. density,
restitution coefficient, etc. The most realistic case is the
non-Gaussian one, which is common to all experiments and simulations,
while the Gaussian case is presented for reference (after the study of
Martin and Piasecki~\cite{martin}). In the Gaussian case we get
\begin{equation}
\log\frac{K(\mathbf{v},\mathbf{v}')}{K(\mathbf{v}',\mathbf{v})}=\frac{\Delta}{2T} +2 \frac{1-\alpha}{1+\alpha}\frac{\Delta}{2T}=\frac{3-\alpha}{1+\alpha}\frac{\Delta}{2T},
\end{equation}
with $\Delta=v_{\sigma}^2-(v_{\sigma}')^2 \equiv |v|^2-|v'|^2$,
i.e. the kinetic energy lost by the test-particle during one
collision.  In the First Sonine correction case it is instead obtained
\begin{equation} \label{ratio_sonine}
\log\frac{K(\mathbf{v},\mathbf{v}')}{K(\mathbf{v}',\mathbf{v})}=\frac{3-\alpha}{1+\alpha}\frac{\Delta}{2T}+\log\frac{\left\{1+a_2S_2^{d=1}\left[\frac{\left(\frac{2}{1+\alpha}(v_{\sigma}'-v_{\sigma})+v_{\sigma}\right)^2}{2T}\right]\right\}}{\left\{1+a_2S_2^{d=1}\left[\frac{\left(\frac{2}{1+\alpha}(v_{\sigma}-v_{\sigma}')+v_{\sigma}'\right)^2}{2T}\right]\right\}}.
\end{equation}

In the case where $P(v)$ is a Gaussian with temperature $T$, it is immediate
to observe that
\begin{equation}
P_*(\mathbf{v})K(\mathbf{v},\mathbf{v}')=
P_*(\mathbf{v}')K(\mathbf{v}',\mathbf{v})
\end{equation}
if $P_*$ is equal to a Gaussian with temperature
$T'=\frac{\alpha+1}{3-\alpha}T \le T$. This means that there is a Gaussian
stationary solution of equation~\eqref{eq:markov} (in the Gaussian-bulk case),
which satisfies detailed balance. The fact that such a Gaussian with a
different temperature $T'$ is an exact stationary solution was known
from~\cite{martin}. It thus turns out that detailed balance is
satisfied in these cases, even in the absence of thermal equilibrium. 
Of course this is an artifact of our model: it is indeed highly unrealistic 
that a granular gas yields a Gaussian velocity pdf. As soon as the gas velocity 
pdf $P(v)$ ceases to be Gaussian,
detailed balance is violated, i.e. the stationary process performed by the
tracer particle is no more in equilibrium within the thermostatting gas.  

\subsection{Numerical results}

\begin{figure}[htbp]
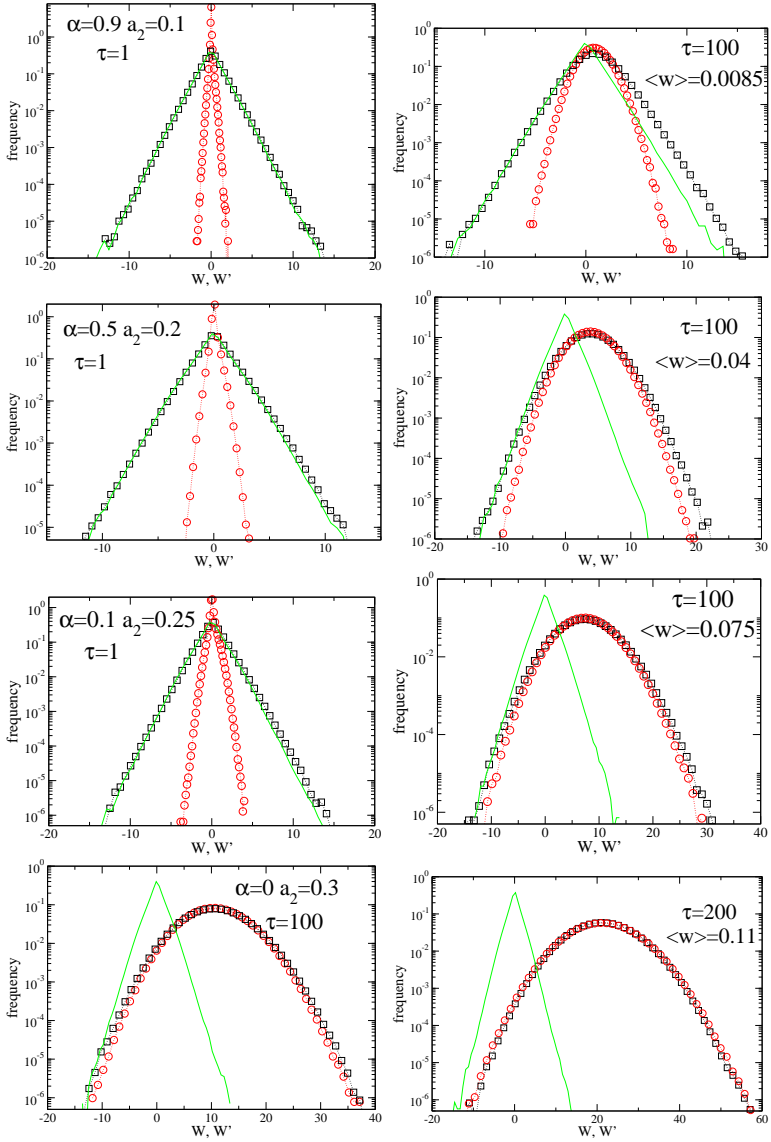

  \includegraphics[width=5cm,clip=true]{pdf_a09_s01_t1.eps}
  \includegraphics[width=5cm,clip=true]{pdf_a09_s01_t100.eps}\\
  \includegraphics[width=5cm,clip=true]{pdf_a05_s02_t1.eps}
  \includegraphics[width=5cm,clip=true]{pdf_a05_s02_t100.eps}\\
  \includegraphics[width=5cm,clip=true]{pdf_a01_s025_t1.eps}
  \includegraphics[width=5cm,clip=true]{pdf_a01_s025_t100.eps}\\
  \includegraphics[width=5cm,clip=true]{pdf_a0_s03_t100.eps}
  \includegraphics[width=5cm,clip=true]{pdf_a0_s03_t200.eps}
\caption{{\bf Color online}. Probability of fluctuations of $W(\tau)$
  (black squares), $\oW(\tau)$ (red circles) and $B=\oW-W$ (green
  line) for the continuous Markov process defined by the dynamics of a
  tracer in a granular gas. Each line is composed of two graphs and
  shows the results for a particular choice of $\alpha$ and $a_2$ (the
  coefficient of the first Sonine correction characterizing the
  non-Gaussianity of the velocity pdf of the gas): the left graphs are
  at small times, while the right graphs are at large times. Mean free
  time between collisions slightly varies with parameters, but is
  always near $\tau_c \sim 0.3$.  For each choice of parameters,
  inside the right graph, it is shown the value of $\langle W(\tau)
  \rangle$ and of $\langle w \rangle \equiv \langle W(\tau)
  \rangle/\tau \equiv \langle \oW(\tau) \rangle/\tau \equiv \lim_{\tau
  \to \infty} W(\tau)/\tau \equiv \lim_{\tau \to \infty}
  \oW(\tau)/\tau$ which can be considered a distance from
  equilibrium. \label{fig:pdf}}.
\end{figure}

\begin{figure}[htbp]
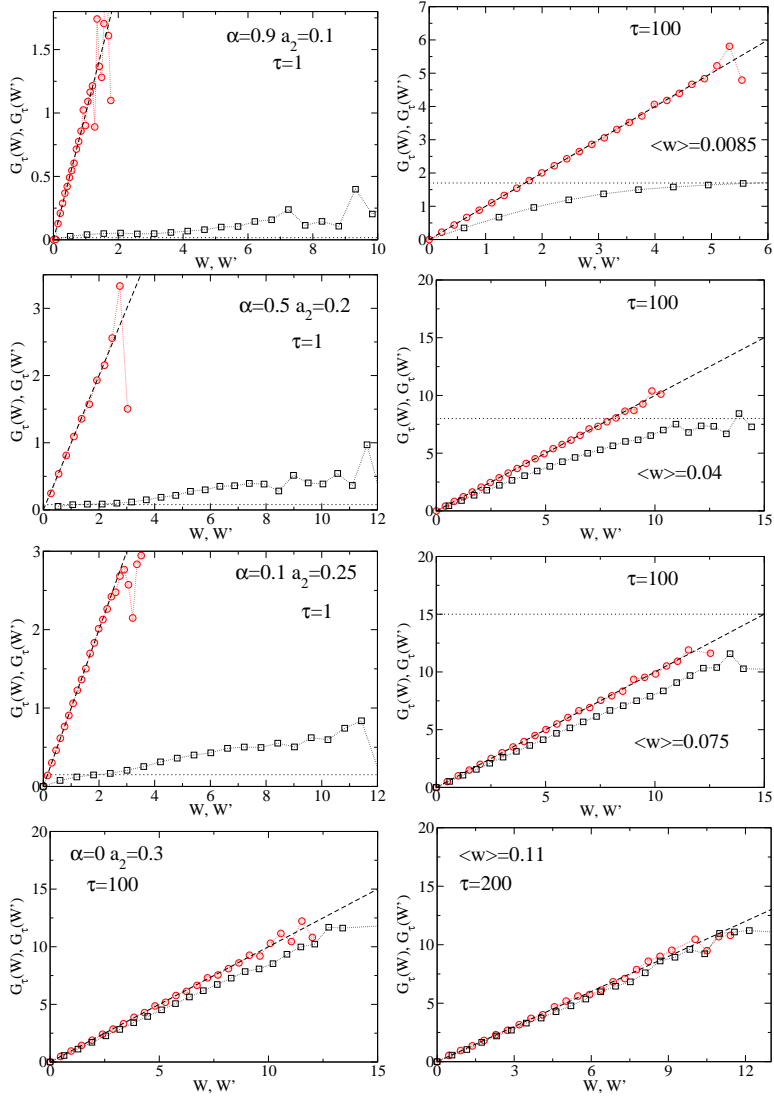

  \includegraphics[width=5cm,clip=true]{gc_a09_s01_t1.eps}
  \includegraphics[width=5cm,clip=true]{gc_a09_s01_t100.eps}\\
  \includegraphics[width=5cm,clip=true]{gc_a05_s02_t1.eps}
  \includegraphics[width=5cm,clip=true]{gc_a05_s02_t100.eps}\\
  \includegraphics[width=5cm,clip=true]{gc_a01_s025_t1.eps}
  \includegraphics[width=5cm,clip=true]{gc_a01_s025_t100.eps}\\
  \includegraphics[width=5cm,clip=true]{gc_a0_s03_t100.eps}
  \includegraphics[width=5cm,clip=true]{gc_a0_s03_t200.eps}
\caption{{\bf Color online}.  $G_\tau(W)$ vs. $W$ (black squares) and
  $G_\tau(\oW)$ vs. $\oW$ (red circles) for the continuous Markov
  process defined by the dynamics of a tracer in a granular gas. Each
  line is composed of two graphs and shows the results for a
  particular choice of $\alpha$ and $a_2$ (the coefficient of the
  first Sonine correction characterizing the non-Gaussianity of the
  velocity pdf of the gas): the left graphs is at small times and the
  right graphs at large times. For each choice of parameters, inside
  the right graph, it is shown the value of $\langle W(\tau) \rangle$
  and of $\langle w \rangle \equiv \langle W(\tau) \rangle/\tau \equiv
  \langle \oW(\tau) \rangle/\tau \equiv \lim_{\tau \to \infty}
  W(\tau)/\tau \equiv \lim_{\tau \to \infty} \oW(\tau)/\tau$ which can
  be considered a distance from equilibrium. The dashed line has slope
  $1$. The dotted line represents the van Zon and
  Cohen prediction for the $FR_W$ violation, $G_\tau(W)=2 \langle W
  \rangle$ for large $W$~\cite{vanzon}. It is not visible in the
  fourth line. \label{fig:gc}}
\end{figure}

The dynamics of a tracer particle undergoing inelastic collisions with
a gas of particles in a stationary state with a given velocity pdf
$P(\mathbf{v})$ has been simulated by means of the Direct Simulation
Monte Carlo algorithm~\cite{bird}.  We have three parameters: the
restitution coefficient $\alpha$ of collisions between the tracer and
the rest of the gas, the temperature of the gas $T$ (which we take as
unity) and the coefficient of the first Sonine correction which
parametrizes the velocity pdf of the gas, $a_2$. The tracer particle
has a measured ``temperature'' (mean kinetic energy) $T_*<T$ and a
velocity pdf that is observed to be well described again by a first
Sonine correction to a Gaussian, parametrized by a coefficient
$a_2^*$.  To measure the quantity $\oW(\tau)=W(\tau)+B$ on each segment of
trajectory we assume this observation to be exact and we use the
values $T_*$ and $a_2^*$ measured during the simulation itself to
compute the ``boundary term'' $B$. All the results are shown in
figures~\ref{fig:pdf} and~\ref{fig:gc}, for two different choices of the time $\tau$
and many choices of the parameters $\alpha$ and $a_2$. We display in
each figure the value of $\langle w \rangle=\lim_{\tau \to \infty}
W(\tau)/\tau$. Again the idea that this quantity measures the distance from
equilibrium of our system is well supported by the results of the
simulations: $\langle w \rangle$ is zero when $a_2=0$ and increases as
$\alpha$ is decreased and $a_2$ is increased. We recall that
$\lim_{\tau \to \infty} W(\tau)/\tau=\lim_{\tau \to \infty}
\oW(\tau)/\tau\equiv \langle w \rangle$, since the difference between
the two functionals has zero average at large times $\tau$. We also
remark that the distribution of both quantities $W$ and $\oW$ are
symmetric at equilibrium (i.e. when $a_2=0$).

We first discuss the results concerning the fluctuations of $W(\tau)$,
identified in figures~\ref{fig:pdf} and~\ref{fig:gc} by squared
symbols.  They are strongly non-Gaussian, with almost exponential
tails, at low values of $\tau$ for any choice of the parameters.  At
large values of $\tau$ (many hundreds of mean free times), the
situation changes with how far from equilibrium the system is. At low
values of $\langle w \rangle$ the tails of the distribution are very
similar to the ones observed at small times. At higher values of
$\langle w \rangle$ the distribution changes with time and tends to
become more and more Gaussian. We have not been able to measure
negative deviations larger than the ones shown in the
figures. Distributions obtained at higher values of $\tau$ yielded a
smaller statistics and very few negative events. With these numerical
observations, we could not verify the $FR_W$ relation.  In the last
case, which is very far from equilibrium ($a_2=0.3$, $\alpha=0$), a
behavior compatible with $FR_W$ is observed, i.e. $G_\tau(W) \approx
W$, but the range of available values of $W$ is much less than
$\langle W \rangle$. At this stage, and in practice, we consider such
results a failure of the $FR_W$ for continuous Markov processes. Our
numerical results are also in fair agreement with the prediction by
van Zon and Cohen for the violation of the $FR_W$, i.e. (at least for
large times and not too far from equilibrium) $G_\tau(W) \sim 2
\langle W \rangle$. The deviations from the Gaussian of the pdfs of
$\oW$ are weaker than those observed for $W$, and the $FR_{\oW}$ is
well satisfied in all non-equilibrium cases at all times $\tau$. The
verification of both FRs is either trivial or meaningless in the
equilibrium cases ($a_2=0$), where $G_\tau \equiv 0$.

\section{Conclusions}

We conclude this paper with the following observations.  In the first
place, it seems important to point out the remarkable similarity of
the results obtained for a rather abstract Markov chain and the
granular tracer. Because the observables considered in this paper are
subtle, one may expect that the similarity between the abstract and
the concrete models concerns other observables as well, which is worth
further investigation.

We have observed a failure of the fluctuation relation for $W$
($FR_W$), at all times that we could consider. This does not exclude
that the relation holds at much larger times, which are of no
practical interest, although they can be of theoretical interest.  The
failure is due to the boundary terms, which become more and more
important as the steady state draws closer and closer to
equilibrium. This suggests that the difference between physically
relevant times, and the times at which the $FR_W$ might be verified,
grows as the steady state approaches an equilibrium state, possibly
diverging in the equilibrium limit.

The (practical) failure of the $FR_W$ reported here does agree in a
quantitative fashion with the one predicted by Van Zon and Cohen,
which was thought to be at work also in some deterministic
systems~\cite{galla_last}.  However, a
different conclusion is afforded by the results of~\cite{harris},
which indicates that the possible failures of the $FR_W$ do not have a
universal character, and that further investigations are desirable.

%
%

The general conclusion that we can draw from the above observations is
that large deviations are particularly subtle objects, whose
peculiarities are not so obvious in the physics of equilibrium systems,
but become evident as soon as a slight perturbation away from
equilibrium sets in. This should not sound paradoxical: steady states
are achieved asymptotically in time; this allows phenomena, such as
long range correlations, which are extraneous to the
physics of most equilibrium systems, but are well known in the physics
of nonequilibrium steady states. Furthemore, to keep a system on a
nonequilibrium steady state, an unlimited amount of work may be done
on the system or extracted from it, and this is bound to have
observable (and perhaps exploitable) consequences on the behaviour of
the system.


\noindent {\it Acknowledgments.--} A. P. acknowledges the support of
the EC grant MERG-021847. The authors are indebted to D. J. Searles
for discussions, hospitality and a critical reading of the
manuscript. A. P. also acknowledges discussions with G. Gallavotti,
P. Visco, F. van Wijland, and F. Zamponi.


\end{document}